\documentclass[aps,preprint,superscriptaddress,showpacs,showkeys,nofootinbib,12pt]{revtex4}
\usepackage{graphicx}

\usepackage[english]{babel}

\newcommand{\be}{\begin{eqnarray}}
\newcommand{\ee}{\end{eqnarray}}
\newcommand{\bea}{\begin{eqnarray}}
\newcommand{\eea}{\end{eqnarray}}

\begin{document}
\title
{\bf Newtonian motion as origin of anisotropy of the local velocity field of
galaxies}

\author{Monika Biernacka}
\affiliation{Pedagogical University, Institute of
Physics, 25-406 Kielce, Poland}

\author{Piotr  Flin}
\email{sfflin@cyf-kr.edu.pl} \affiliation{Pedagogical University, Institute of
Physics, 25-406 Kielce, Poland} \affiliation{Bogoliubov Laboratory of
Theoretical Physics, Joint Institute for Nuclear Research, 141980, Dubna,
Russia}

\author{Victor Pervushin}
\email{pervush@thsun1.jinr.ru} \homepage[Home page:
]{http://thsun1.jinr.ru/~pervush/} \affiliation{Bogoliubov Laboratory of
Theoretical Physics, Joint Institute for Nuclear Research, 141980, Dubna,
Russia}

\author{Andrey Zorin}
\affiliation{Faculty of Physics, MSU, Vorobjovy Gory, Moscow,
119899, Russia}

 \begin{abstract}
The origins of recently reported anisotropy of the local velocity
field of nearby galaxies (velocities $<$ 500 km/s corresponding to
the distance less than 8 Mpc) are studied. The exact solution of
the Newtonian equation for the expanding Universe is obtained.
This solution allows us to separate the Newtonian motion of nearby
galaxies from the Hubble flow by the transition to the conformal
coordinates. The relation between the Hubble flow and the
Newtonian motion is established. We show that the anisotropic
local velocity field of nearby galaxies can be formed by such a
Newtonian motion in the expanding Universe, if at the moment of
the capture of galaxies by the central gravitational field.
\end{abstract}






\maketitle

\section{Introduction }%

Recent observation of the local velocity field of galaxies gives a
three-dimensional ellipsoid with different values of the Hubble
parameter, clearly showing its  anisotropic character
~\cite{Karach1,Karach2}.

  In this paper, we present a possible point of view that this
 local velocity field of galaxies can be explained by their Newtonian
 motions.

 The analysis of the observational data will be based on
 the radial velocities of nearby galaxies, belonging to the Local
 Group. Our paper is organized in the following manner. In Section II,
 the cosmic evolution is described.  In Section III, we introduce the Newtonian
 motion and separate this motion from the cosmic one.
   In  Section IV, the initial data of the galaxies capture
   by a central gravitational
  field is considered.  In  Sections V, the simplest example
   is given to elucidate our results.
   The paper ends with the conclusions.

\section{Cosmic evolution of a free particle}

Effects of cosmic evolutions are considered
 in the Friedmann--Lem\^{a}itre--Robertson--Walker (FLRW) metrics
\be\label{gr5} (ds^2)=(dt)^2-a(t)^2(dx^i)^2. \ee
 The formulation of the Newtonian problem in this metrics proposes a
 choice of physical variables and coordinates.
 The modern cosmology uses two
 choices of such variables:
 the conformal time $(\eta)$ and coordinate distance $(x^i)$
 with the interval
 \be\label{R}
 (ds^2)={a}(\eta)^2\left[(d\eta)^2-(dx^i)^2\right],
 \ee
 and the Friedmann time $(t)$ and distance $X^i=a x^i$ in terms of
 which the interval (\ref{gr5}) takes the form
  \be\label{A}
 (ds^2)=(dt)^2-\left[dX^i-H(t)X^idt\right]^2,
 \ee
 where $H(t)=\dot a(t)/a(t)$ is the Hubble parameter,
 we have used here the formula of differential
 calculus $adx=d(ax)-xda$.  Both the sets are mathematically equivalent\footnote{Note that $a(t)=a(\eta(t))$
 is connected with $a(\eta)$ by equation $dt=a(\eta)d\eta$.}.

 In terms of the Friedmann variables  $X^i=a x^i$
 the Newton action in the space with the interval (\ref{A}) takes the form
 \be\label{na1}
 S_A=\int\limits_{t_I}^{t_0}dt\left[P_i(\dot X^i-HX^i)-
 \frac{P_i^2}{2m_0} \right].
 \ee
 The equations of motion
 \be\label{na2}
 \dot X^i-HX^i=\frac{P_i}{m_0},~~~~~~~~~\dot P^i+HP^i=0
 \ee
 have the simplest solution $X^i=a x_0^i$, where $x_0^i$ is a constant.

 The transition to the conformal variables gives the action\footnote{The action (\ref{na1})
 and (\ref{na4}) differ from the one (7.4) in monograph
\cite{Peebles}, that is not compatible with quantum field theory
on the conformal flat metric (2) \cite{J:Phys}.}
 \be\label{na4}
 S_A=\int\limits_{\eta_I}^{\eta_0}d\eta\left[p_i\frac{dx^i}{d\eta}-
 \frac{p_i^2}{2m_0a(\eta)} \right]
 \ee
 for a particle with the running mass $m_0a(\eta)$ (with the present--day value
 $a(\eta_0)=a_0=1$). This transition
  is just  the main idea of our paper to
 separate the Hubble velocity field
 \be\label{def:H:tot} H_{\rm tot}=\frac{1}{R}\frac{dR}{dt}
  ~~~~ ~~~~ \left(R=\sqrt{X_1^2+X_2^2+X_3^2}\right) \ee from possible Newtonian motion.
Substituting $R=ar$, where $r=\sqrt{x_1^2+x_2^2+x_3^2}$,
 in the definition of the total Hubble flow (\ref{def:H:tot}) we get
 this Hubble flow in the following form
 \be H_{\rm tot}=\frac 1a \frac
{da}{dt} + \frac 1r \frac {dr}{dt} = H + \Delta H. \ee One can see
that the total Hubble flow differs from the classical one $H=\dot
a/a$  by the value of the local flow \be\label{cr11rv} \Delta
H=\frac {1}{r}\frac{dr}{dt}.
 \ee

\section{Newtonian motion in an expanding Universe}
 Let us consider the action
 \be\label{cr11a}
 S_A=\int\limits_{t_I}^{t_0}dt\left[P_i(\dot X^i-HX^i)-
 \frac{P_i^2}{2m_0} +\frac{\alpha}{R}\right],
 \ee
 where
 $\alpha={M_{\rm O} m_0 G}$ is a
  constant of a Newtonian
 interaction of a galaxy with a mass $m_0$ in a gravitational field of a central mass
  ${M_{\rm O}}$.
Action  (\ref{cr11a}) for radial momentum $P_R$ and orbital moment
$P_\theta$ in the cylindrical coordinates
 \be
 \label{coord}
 X^1=R\cos\theta,~~X^2=R\sin\theta,~~X^3=0
 \ee
 takes the form
 \be\label{cr11ar}
 S_A=\int\limits_{t_I}^{t_0}dt\left[P_R(\dot R-HR)+P_\theta \dot \theta-
 \frac{P_R^2}{2m_0} -\frac{P_\theta^2}{2m_0R^2}+\frac{\alpha}{R}\right].
 \ee

To separate the Newtonian motion from the Hubble velocity field,
we use to the conformal variables $p_r=P_Ra(t),~r=R/a(t),
~d\eta=dt/a(t)$. In terms of these variables the action takes the
form \be\label{cr11r}
 S_A=\int\limits_{\eta_I}^{\eta_0}d\eta\left[p_rr'+P_\theta \theta'-
 \frac{p_r^2}{2m_0{a}(\eta)} -\frac{P_\theta^2}{2m_0{a}(\eta) r^2}+
 \frac{\alpha}{r}\right].
 \ee

\section{The capture of
galaxies by the central gravitational field}
The energy  of a particle with the running mass $m(\eta)=a(\eta)m_0$ described
by the action (\ref{cr11r}) \be \label{ce}
E(\eta)=\frac{p_r^2}{2m_0{a}(\eta)}+\frac{P_\theta^2}{2m_0{a}(\eta)r^2}-
 \frac{\alpha}{r}
\ee is not conserved in the contrast to the energy of particle
with a constant mass in the Newtonian  mechanics \cite{L-L:1}. In
our case (\ref{ce}), if the scale factor $a(\eta)$ increases, the
energy (\ref{ce}) runs from its positive values to negatives ones.
There is a moment of a time $\eta=\eta_I$ when the energy
(\ref{ce}) is equal to zero: \be \label{1ce}
\frac{E(\eta_I)}{m_0{a}(\eta_I)}\equiv
\frac{(r'_I)^2+v_I^2}{2}-w_I^2=0, \ee where
$r'_I=p_r(\eta_I)/m_Ir_I$ is radial initial velocity,
$m_I=m_0a(\eta_I),~r_I=r(\eta_I)$ are the initial conformal mass
and coordinate distance, and \be \label{or} v_I=
\frac{P_\theta}{m_Ir_I}, ~~~~~~~ w_I=\sqrt{\frac{\alpha}{m_Ir_I}}
\ee are the orbital   velocity and Newtonian one, respectively. It
is known that the change of a sign of the energy means the change
of an unrestricted motion of a particle by a finite motion in the
central field. Therefore, the time $\eta_I$ can be treated as the
time of
 the capture of a particle (cosmic object) by the central gravitational field.

If the initial radial velocity is too equal to zero $r'_I=0$, the
zero--energy constraint (\ref{1ce}) \be \label{2ce} v_I^2=2w_I^2
\ee becomes the equation for the initial data
${m_0a(\eta_I)r(\eta_I)}\equiv m_Ir_I$. The solution of this
equation $m_Ir_I=P^2_\theta/(2\alpha)$ can give a orbital velocity
for all trajectories of the captured cosmic objects \be
\label{3ce} v_I=\frac{2\alpha}{P_\theta}=\rm{constant} \ee The
fact of the universality of the orbital velocity (\ref{3ce}) for
all  ellipsoidal trajectories (due to the zero energy initial data
of the formation of a local universe) gives us a possibility of
expanding both the numerous observational data on the law of the
constant orbital velocity \cite{Karach2,L-L:1,ch,E:1} and the
anizotropy of the local velocity field \cite{Karach1,Karach2}.

The anisotropy of the local velocity field  $\Delta H=0$
\cite{Peebles,J:Phys} is not compatible with the class of the
isotropic circular trajectories $r(\eta_0)\equiv r_0, r'=r''=0$
with the equation of motion \be \label{4ce} v_0^2=w_0^2
~~~~~~~~~~~~~~~\left(v_0=\frac{P_\theta}{m_0r_0},~~~
w_0=\sqrt{\frac{\alpha}{m_0r_0}}~\right) \ee known as the ``virial
theorem'' \cite{E:1,E:2,pri}, where the initial radius $r_I$ can
be identified with the observational radius\footnote{Remind that
the  hypothesis of the Cold Dark Matter
 was proposed in \cite{E:1,E:2} to explain
the law of the constant orbital velocity  in the class of the
circular trajectories in the nonexpanding  Universe where
(\ref{4ce}) is valid.}. If trajectories belong to a class of
ellipsoidal ones, the initial radius $r_I$ does not coincide with
the observational radius.

\section{ The exact solution}

Let us consider a solution of the Kepler problem, i.e., a motion
of an object in the central field in the expanding Universe for
the rigid state when the densities of energy and pressure are
equal. The cosmic scale factor $a$ can be written as \cite{039}
 \be\label{hab2}
 a(\eta)^2=[1+2H_0(\eta-\eta_0)],
 \ee
 which describes the recent Supernova data \cite{snov,riess,sn1997ff} in
 the relative units \cite{039}. Equation (\ref{hab2}) can be considered
 as a change of the evolution parameter
 $\eta\to a(\eta)$.
        In terms of the new variables
$$y=r/r_0,~p_y=p_r/{m}_0,~v_0=P_\theta/r_0{m}_0,$$
 where $r(\eta_0)=r_0$ is the present--day data,
  action (\ref{cr11r}) takes the form
 \be\label{cr11r1}
 S_A=r_0m_0\int\limits^{1}_{a_I}da\left\{p_y\frac{dy}{da}+
 v_0\frac{d\theta}{da}-
 \frac{1}{c_0}\left[\frac{p_y^2+v^2_0/y^2}{2} -
 \frac{aw^2_0}{y}\right]\right\},
 \ee
where $w_0=\sqrt{\alpha/m_0r_0}$, $c_0=H_0 r_0$ are the Newtonian velocity and the
Hubble one, respectively, and $a_I=1/(1+z_I)$ is determined with the redshift
$z_I$ at the moment of formation $\eta=\eta_I$. The total energy of the system
(\ref{ce}) in terms of the new variable takes the form \be \frac{E(a)}{r_0m_0}
= \frac{1}{c_0}\left[\frac{p_y^2+v^2_0/y^2}{2} -
 \frac{aw^2_0}{y}\right].\ee

It is easy to see that $v_0$ is a constant of the motion $dv_0/da=0$.
The equations of the radial motion
 \be\label{cr11re}
 p_y&=&c_0\frac{dy}{da}, ~~~~~~~~~ c_0\frac{dp_y}{da}=\frac{v^2_0}{y^3}-\frac{aw^2_0}{y^2} \ee can be
written in the Lagrangian form \be\label{eq:y}
\frac{d^2y}{da^2}&=&\left(\frac{w_0}{c_0}\right)^2
\left[\left(\frac{v_0}{w_0}\right)^2\frac{1}{y^3}-\frac{a}{y^2}\right].
\ee At the present--day time $a=a_0=1,~y_0=1$ this equation
determines the second derivative \be\label{eq:y0}
\left.\left[\frac{d^2y}{da^2}\right]\right|_{a=1}&=&\left(\frac{w_0}{c_0}\right)^2
\left[\left(\frac{v_0}{w_0}\right)^2-1\right] \ee in terms of
three velocities $w_0,v_0,c_0$, and it allows us to choose any
relations between $w_0$ and $v_0$, in particular $v^2_0=2w^2_0$,
in contrast to the circular trajectory where $v^2_0=w^2_0$.

The
general solution of  this equation (\ref{eq:y}) can be obtained in the
parametrical form:
 \be
a (\tau) &=& c_1 {N_2 (\tau)  \over \tau^{2/3}N (\tau) },~~~~~~~~~~~~~~~~~
y(\tau)= c_2\tau^{2/3}N (\tau) ,\label{sol:ra}\ee where $\tau$ is the parameter
of solution with the initial date $r_0=r(\eta_0)$, and \be \!\!\!\! \!\!\!\! N
(\tau)
&=&\alpha_1 U^2 (\tau) +\beta_1 U (\tau)  V (\tau)  + \gamma_1 V ^2 (\tau) , \\
N_2 (\tau) &=& \left(\tau\frac{dN (\tau) }{d\tau}+\frac 23 N (\tau)
\right)^2\pm 4\tau^2N ^2(\tau)   +\omega^2 \Delta, \label{def:N2}\\ \Delta &=&
4 \alpha_1 \gamma_1 - \beta_1^2, \\c_1&=&\left(\frac{v_0}{w_0}\right)
\left(\frac{3c_0}{4w_0}\right)^{1/3}\frac{1}{2\omega\Delta^{1/2}},\\
c_2&=&\left(\frac{v_0}{w_0}\right)
\left(\frac{4w_0}{3c_0}\right)^{1/3}\frac{1}{\omega\Delta^{1/2}}.\ee
 Three constants $ \alpha_1,~\beta_1,
~\gamma_1={\rm const}$ can be found from the following system of
three equations:
\be\left.a\right|_{\tau=0}=\left.a\right|_{\eta=\eta_I}=a_I=\frac{1}{1+z_I},
~~~~~ \left.y\right|_{\tau=0}=\frac{r_I}{r_0}=1+z_I, ~~~~~
\left.\frac{dy}{da}\right|_{\tau=0}=0 ; \ee here for the upper
sign (restricted solution at infinity, $\tau=+\infty$) in
(\ref{def:N2}) \be\label{solution:restricted} U (\tau)
=J_{1/3}(\tau),~~~~V (\tau) =Y_{1/3}(\tau),~~~~{\rm and}~~~~
\omega=\frac 2\pi,\ee where $J_{1/3}(\tau)$ and $Y_{1/3}(\tau)$
are the Bessel functions of the first and second (or Niemann
function) kind, while for the lower sign (unrestricted solution at
infinity, $\tau=+\infty$) in (\ref{def:N2})
\be\label{solution:unrestricted} U (\tau) =I_{1/3}(\tau),~~~~V
(\tau) =K_{1/3}(\tau),~~~~{\rm and}~~~~ \omega=-1,\ee where
$I_{1/3}(\tau)$ and $K_{1/3}(\tau)$ are the modified Bessel
functions of the first
 and second (or MacDonald function)  kind (see, e.g. \cite{ZP}).

\begin{figure}[htbp]
\begin{center}
\begin{minipage}[t]{.48\textwidth}
  \centering
\includegraphics[width=\textwidth]{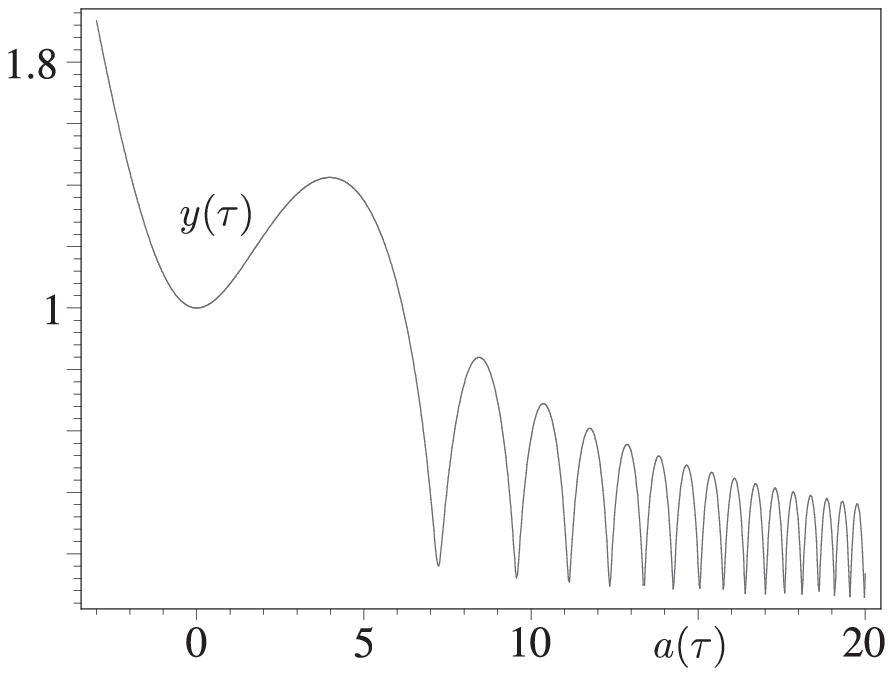}
\vspace{-5mm} \caption{\small Graph of function $y(\tau)=a(\tau)r(\tau)/r_0$
(\ref{sol:ra}) at $\displaystyle v_0^2/c_0^2=0.2$ and $\displaystyle
w_0^2/c_0^2=0.1$.}\label{fig:y:a}
\end{minipage}
\hskip.02\textwidth
\begin{minipage}[t]{.48\textwidth}
  \centering
\includegraphics[width=\textwidth]{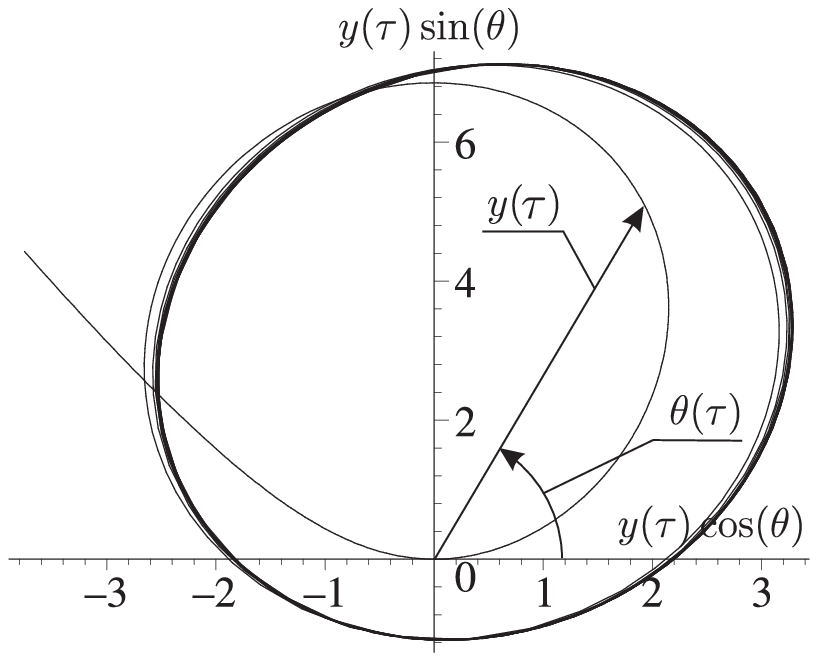}
\vspace{-5mm} \caption{\small Graph of functions product $a(\tau)y(\tau)$
(\ref{sol:ra})  at $\displaystyle v_0^2/c_0^2=0.2$ and $\displaystyle
w_0^2/c_0^2=0.1$ for $-3<a(\tau)<14$.}\label{fig:y:theta}
\end{minipage}
\end{center}
\end{figure}

 A solution of the equation of motion
following from the action (\ref{cr11r}) is given in Fig. \ref{fig:y:a} and
Fig. \ref{fig:y:theta}. We use this solution for construction of two plots.

\begin{figure}[htbp]
\begin{center}

\begin{minipage}[t]{.48\textwidth}
  \centering
\includegraphics[width=\textwidth]{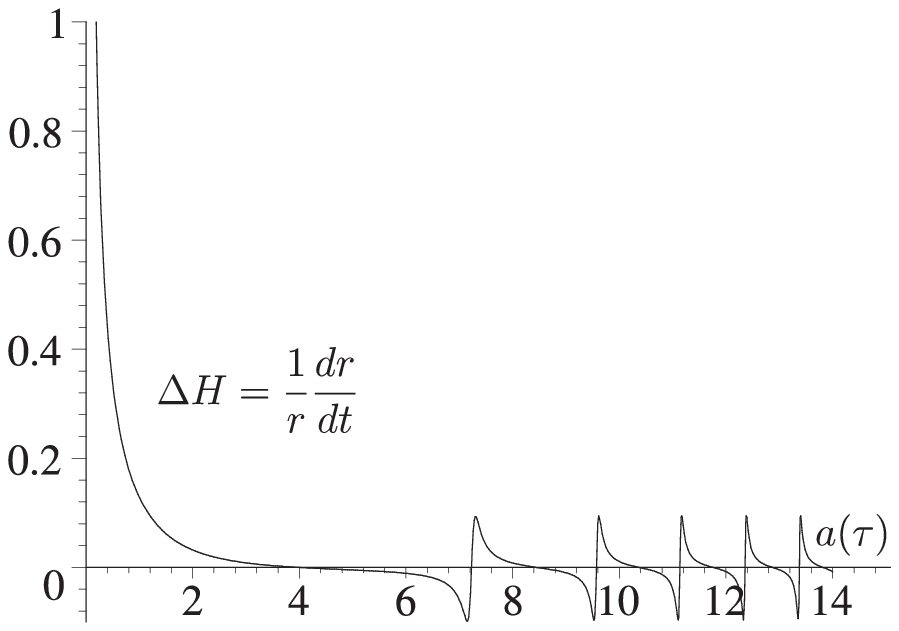}
\vspace{-5mm} \caption{\small Graph of function $\Delta H$ (\ref{cr11rv}) in
units $H_0$ at $\displaystyle v_0^2/c_0^2=0.2$ and $\displaystyle
w_0^2/c_0^2=0.1$.}\label{fig:DH}
\end{minipage}
\hskip.02\textwidth
\begin{minipage}[t]{.48\textwidth}
  \centering
\includegraphics[width=\textwidth]{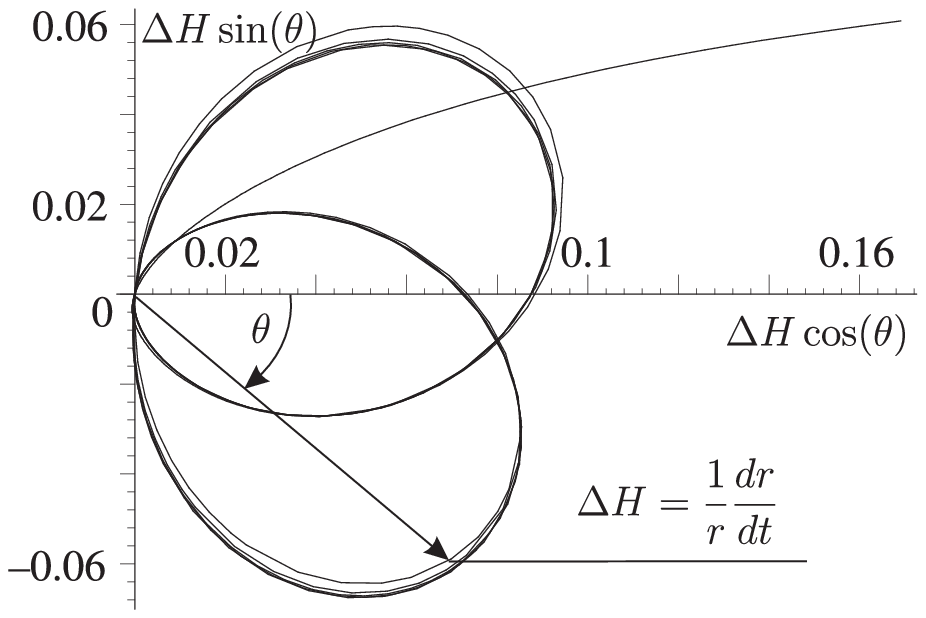}
\vspace{-5mm} \caption{\small Graph of function $\Delta H$ (\ref{cr11rv}) in
units $H_0$ at $\displaystyle v_0^2/c_0^2=0.2$ and $\displaystyle
w_0^2/c_0^2=0.1$ for $0.8<a(\tau)<14$.}\label{fig:DH:theta}
\end{minipage}
\end{center}
\end{figure}

Figure \ref{fig:DH} gives the values of the correction (\ref{cr11rv}) to the
Friedmann Hubble flow resulting from taking into account eq.
(\ref{cr11rv}). It is clearly seen that the corrections are dumped with
time and have a quasi--periodical character.

In Fig. \ref{fig:DH:theta} the angular distribution of the Hubble flow
correction, as given by (\ref{cr11rv}), is presented. The correction is
anisotropic. We consider the 2-dimensional case while Karachentsev's anisotropy
is observed in 3-dimensions. Nevertheless, our 2-dimensional analysis allows
one to see the anisotropy of the Hubble flow and to estimate the order of the
magnitude of the anisotropy.

We have considered the case when the formation of a galaxy began
from the zero--energy state (the initial data $E_0=0$, and
velocity $y_0'=0 $). These data correspond to the relation
$v_0^2=2w_0^2$.

\section{Conclusion}
Our paper was motivated by the finding that in the local Universe
the velocity field is anisotropic \cite{Karach1,Karach2}. This
effect is difficult for explanation. The only possible suggestion,
but rejected by Karachentsev, was rotation \cite{Karach2}. We are
trying to find the origin of this anisotropy. In ordered to do
this, we consider the general uniform expansion of the Universe.
Since this is the nearby (less than 8 Mpc) part of the Universe
around us,
 we use
the Newtonian approach. We studied the motion of the test massive
particle in the central gravitational field on the background of
cosmic evolution of the type of FLRW space--time with uniform
expansion. We assume the rigid state of the matter when densities
of energy and pressure are equal and which corresponds to
conformal cosmology \cite{039} compatibles with Supernova data
\cite{snov}. We obtained the exact solution of the
above--mentioned Kepler problem, to find the difference between
the uniform Hubble flow and our case.
 We have shown that this difference was anisotropic. In such a way we
explained the anisotropy of the local velocity field by
the Newtonian motion of
galaxies in the central field. Of course, our 2--D consideration  shows a
possible mechanism of the observed 3--D anisotropy.

Having the solution of the Kepler problem we admit the rotation of
galaxies around the center of the local Universe. This local
Universe must be regarded as the Local Group of galaxies. In such
a way, we support the picture in which galaxies rotate around the
center of the Local Group in the class of ellipsoidal
trajectories.


\newpage

\end{document}